\documentclass{article}
\usepackage{spconf,amsmath,amssymb,graphicx,booktabs,url}
\usepackage[hidelinks]{hyperref}
\usepackage{xurl}

\title{Multi-Rate Bandwidth Extension by Token Completion\\ in Neural Audio Codecs}

\name{Beno\^{i}t Ginies, Olivier Fercoq, Ga\"el Richard\thanks{This work was funded by the European Union (ERC, HI-Audio, 101052978). Views and opinions expressed are those of the authors only and do not necessarily reflect those of the European Union or the European Research Council.}}
\address{LTCI, T\'el\'ecom Paris, Institut Polytechnique de Paris, Palaiseau, France}

\begin{document}
\ninept
\maketitle

\begin{abstract}
Bandwidth extension aims to reconstruct the high-frequency content missing from a band-limited signal. In this work, we focus on signals processed by a neural audio codec, where the decoder receives band-limited codes and already retains the low-frequency band. The missing content is confined to frequencies above the input's cutoff, transforming bandwidth extension into a token completion task. 
The codec ingests a fixed 48~kHz representation, regardless of the input bandwidth. Thus, the predictor is the only rate-dependent component, and a single model supports input rates of 8, 16, 24, and 32~kHz without explicit bandwidth conditioning, as the latter is recoverable from the codes themselves.
Applied to both a waveform-domain codec (DAC) and a spectral-domain codec (SpectroStream), this method outperforms, on objective metrics, two recent systems on the MUSDB18 dataset across all rates, while operating on a 12~kbit/s token stream. The subjective evaluation of the 8~kHz input rate further shows it performs comparably to the stronger of the two baseline systems.
\end{abstract}


% ===========================================================================
\section{Introduction}
\label{sec:intro}
% ===========================================================================

Bandwidth extension (BWE) reconstructs the high-frequency content of a signal from its band-limited observation. Classical systems replicated or rescaled low-frequency spectra into the missing bands~\cite{Dietz2002SpectralBR}. Neural systems have since reshaped the problem, evolving from early convolutional regressors~\cite{kuleshov2017audio} to more efficient diffusion and flow-matching models~\cite{han2022nu, liu2024audiosr, choi2026universr, kong2025a2sb}, as well as hybrid differentiable-DSP approaches~\cite{grumiaux2023efficient}. Many of these systems support multiple input rates~\cite{han2022nu,liu2024audiosr,choi2026universr}, achieving flexibility by exposing a generative model to the raw band-limited signal. 
This raises the question of whether the same flexibility can be achieved at a lower level, inside a codec, using discrete tokens alone.

Neural audio codecs~\cite{zeghidour2021soundstream, defossez2022high, kumar2023high} convert waveforms into short sequences of discrete tokens. These tokens form a vocabulary that transformers can model~\cite{wang2023neural, copet2023simple}. 
Restoration techniques have similarly emerged in this space: MaskSR~\cite{li2024masksr} and Genhancer~\cite{yang2024genhancer} predict clean codec tokens from degraded inputs for speech enhancement, and discrete diffusion has been adapted to speech BWE~\cite{fang2025vector}. In these pipelines, the decoder operates on discrete codes rather than raw waveforms, allowing restoration modules to process the tokens directly. Also, as these modules are integrated at the decoder, they introduce no additional bitrate overhead.

In our own prior work~\cite{ginies2025soft,ginies2025harmonic} we adopted this approach for music BWE by restructuring the codec itself into soft semantic sections (e.g.\ frequency-dependent sections, or sections separating harmonic and percussive content) which led to improved predictability of the high-branch tokens from the low-branch tokens.  
However, this method inherits a structural limitation: each branch is itself a codec running at its own sampling rate, with no direct capability to accept other input rates. 

This work pursues the same goal from a different perspective. Rather than predicting the full-band signal, the BWE decoder only predicts the content above the input's Nyquist frequency. We freeze the codec and train a single decoder-side transformer to map the codes of a band-limited signal to those of its full-band counterpart. Preserving the true low-frequency content makes the task strictly easier. Moreover, the frozen codec provides a common 48~kHz representation across input bandwidths, making the transformer the only rate-dependent component and enabling a single model to support multiple input rates. Thus, BWE no longer requires codec retraining, but only a single transformer trained across sampling rates. Our contributions are:

\begin{enumerate}\itemsep1pt
\item \textbf{A single predictor covering four input rates.} One transformer, conditioned on nothing but the codec tokens, handles 8/16/24/32~kHz inputs. The cutoff frequency is not supplied, the bandwidth being already recoverable from the codes.
\item \textbf{Competitive quality across two distinct codecs.} On music mixtures, our method outperforms A2SB~\cite{kong2025a2sb} and UniverSR~\cite{choi2026universr} at every input rate. Out of domain, it leads at all rates except the narrowest, 8~kHz. A MUSHRA test at 8~kHz finds our DAC predictor performs comparably to A2SB, the stronger of the two.
The approach generalizes seamlessly from a spectral-domain codec (SpectroStream~\cite{li2025spectrostream}) to a waveform-domain one (DAC~\cite{kumar2023high}), demonstrating that it does not depend on tokens encoding a frequency axis.
\end{enumerate} 
Audio examples are on our website\footnote{\url{https://multi-rate-bwe-by-token-completion.github.io/}} and the implementation is publicly available.\footnote{\url{https://github.com/Multi-Rate-BWE-by-Token-Completion/Multi-rate-BWE}}

\begin{figure*}[t]
\centering
\includegraphics[trim={0.7cm 0.4cm 0.6cm 0.35cm},clip,width=0.65\linewidth]{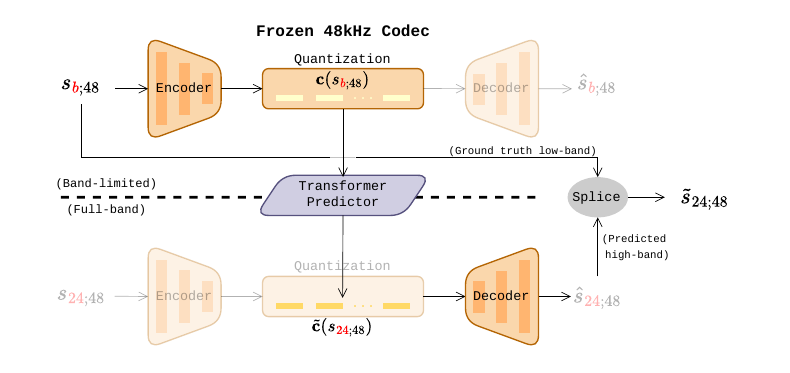}
\caption{Bandwidth extension as token completion. The codec is frozen, so it ingests the same 48~kHz representation whatever the input bandwidth $b$. The transformer completes band-limited codes into full-band ones autoregressively over quantizer depth. Only the first predicted level $\tilde{\mathbf{c}}_1$ is decoded, and its content is spliced above the cutoff onto the ground-truth low band, which reaches the output unchanged.}
\label{fig:pipeline}
\vspace{-10pt}
\end{figure*}

% ===========================================================================
\section{Bandwidth extension as token completion}
\label{sec:method}
% ===========================================================================

In this work, we introduce a decoder-side module that recovers, from the codes of a band-limited signal, the audio the codec would have produced from its full-band counterpart. Two constraints apply throughout: the codec is left unmodified, and no side information is transmitted. Figure~\ref{fig:pipeline} summarizes the resulting system.

\subsection{Task}
\label{ssec:task}

We write $s_{b;SR}$ for a signal band-limited to $b$~kHz and sampled at $SR$~kHz, and $\tilde{s}$ for an estimate obtained by prediction. The target is $s_{24;48}$, and the observation is $s_{b;2b}$, with $2b \in \{8,16,24,32\}$~kHz, resampled before encoding to 48~kHz to give $s_{b;48}$, so that every input reaches the codec at the same sampling rate and no part of the codec is rate-specific.

A frozen codec encodes $s_{b;48}$ into $N_q$ token streams $\mathbf{c}(s_{b;48}) = (c_1,\dots,c_{N_q})$ of $T$ tokens each. The task is to predict $\mathbf{c}(s_{24;48})$ and to use it to synthesize only the band above $b$~kHz, giving $\tilde{s}_{24;48}$. Nothing else is transmitted: the predictor is a decoder-side module, and the bit rate is that of $\mathbf{c}(s_{b;48})$.

\subsection{Application to two different codecs}
\label{ssec:codecs}

The framing assumes nothing about how the codec produces its tokens, so we instantiate it on two codecs that differ in exactly that respect. \textbf{DAC}~\cite{kumar2023high} is waveform-domain: strided 1-D convolutions over the signal give 93.75~Hz frames quantized by a residual vector quantizer (RVQ) of $N_q=12$ codebooks of $1024$ entries, i.e.\ 11.25~kbit/s, with no frequency axis anywhere in the model. \textbf{SpectroStream}~\cite{li2025spectrostream} is spectral-domain: 2-D convolutions run over the complex STFT of 48~kHz audio, downsampling both frequency and time, so a token is attached to a time-frequency region rather than to a stretch of waveform. Its RVQ has $64$ levels of $1024$ entries at 25~Hz, out of which we only use $N_q=48$, i.e.\ 12~kbit/s. No public implementation exists, so we reimplemented it and release it with the predictor. Trained to 500k steps, it reaches ViSQOL $4.46$ at 12~kbit/s against $3.92$ for DAC. Both are retrained at 48~kHz on the data of Sec.~\ref{ssec:data} and then frozen. Whether a token carries an explicit notion of frequency is the one property separating them, and this is further discussed in Sec.~\ref{ssec:domains}.

\subsection{Predictor}
\label{ssec:predictor}

The predictor is a Transformer decoder~\cite{vaswani2017attention} autoregressive over RVQ depth rather than time: at step $j$ it predicts the full-band code $\tilde{c}_j$ at all $T$ time positions simultaneously, conditioned on the $N_q$ band-limited levels and on the $j-1$ full-band levels already predicted. It has 6 layers, $d_{\text{model}}=1024$ and 8 heads. The $N_q$ input streams and the $N_q-1$ feedback streams are embedded at dimension 64, concatenated and projected, and the pooled embedding is layer-normalized. This generalizes the summed embeddings used in~\cite{ginies2025harmonic}, which is the case where the projection is a tiled identity. A learned step embedding tells the network which RVQ level it is predicting.

The input bandwidth is never supplied: the predictor sees codec tokens and nothing else. The model we propose is trained across all four rates at once, $2b$ being drawn uniformly per batch. Sec.~\ref{ssec:design} compares it with four predictors trained one per rate and with a variant that receives a learned embedding of the cutoff. Training is teacher-forced, each step being conditioned on the ground-truth levels below it rather than on the ones it predicted itself, with a cross-entropy loss summed over steps, optimized with AdamW at $10^{-4}$ on a cosine schedule, with a batch size of $16$, for 25k steps.

\subsection{Synthesis protocol}
\label{ssec:synth}

The output $\tilde{s}_{24;48}$ is assembled in the spectral domain from the ground-truth low band of $s_{b;48}$ below $b$~kHz and the decoded prediction above it, the two joined at the cutoff by a 1~kHz crossfade, a width selected on mel distance against 1.5 and 2~kHz. We call this operation the splice, and apply it identically to every system compared, including the baselines of Sec.~\ref{ssec:baselines}.

Decoding uses argmax, and only reconstructs from the first predicted level: we observed that injecting deeper predicted levels did not improve ViSQOL. The predictor is nonetheless trained on all $N_q$ levels, and decoded from its first level by truncating the predicted levels at synthesis. None of this changes the transmitted rate, which remains $N_q$ levels.

% ===========================================================================
\section{Experimental setup}
\label{sec:setup}
% ===========================================================================

\subsection{Data}
\label{ssec:data}

Both codecs are trained on Jamendo~\cite{bogdanov2019mtg} and the MUSDB18 training split~\cite{musdb18}, every file of the two datasets being drawn with equal probability. The predictors are trained on a mixture assembled to carry as much high-frequency energy as possible without shifting the domain too far from that of the codecs: Jamendo, MedleyDB~\cite{bittner2014medleydb}, the MUSDB18 training split and ENST-Drums~\cite{gillet2006enst}, weighted $0.766/0.106/0.064/0.064$, identically for every trained BWE predictor. Jamendo is filtered before entering this mixture, keeping the 39\% of files whose effective bandwidth reaches at least 19~kHz.

The primary test set is 1000 excerpts of 2.5~s from the MUSDB18 test set. The out-of-domain set is 1000 excerpts from OrchideaSOL~\cite{cella2020orchideasol} (solo instruments), a corpus that appears in no baseline's training data. Band-limited partners are built, at test time as in training, by resampling down and back up, which leaves the resampling filter's finite transition band in the signal rather than the sharp edge an ideal low-pass would give.

Neither test set is neutral, which is why both are reported: UniverSR was trained on MUSDB18 without specifying the split, whereas A2SB never saw it. OrchideaSOL is out-of-training for all three systems but sits closer to UniverSR's solo-instrument domain than to our music-mixture one.

\subsection{Metrics}
\label{ssec:metrics}

ViSQOL-audio~\cite{chinen2020visqol} is our main objective metric. We also compute multi-resolution mel and STFT distances~\cite{kumar2023high}, waveform $\ell_1$ and SI-SDR~\cite{le2019sdr}, but SI-SDR is reported rather than used to discriminate between models, being sensitive to sample-level misalignment~\cite{defossez2024moshi}, which matters when the band under evaluation is synthesized rather than reconstructed.

\subsection{Baselines and listening test}
\label{ssec:baselines}

\textbf{A2SB}~\cite{kong2025a2sb} is a diffusion Schr\"odinger bridge over spectrograms at 44.1~kHz. We use the released two-checkpoint ensemble with 50 sampling steps, as additional steps did not improve ViSQOL. Its 44.1~kHz ceiling is not limiting, as our references contain essentially no energy above 22.05~kHz. \textbf{UniverSR}~\cite{choi2026universr} is a vocoder-free conditional flow-matching model operating at 48~kHz and supporting $\{8,12,16,24\}$~kHz inputs, but not 32~kHz. We set its guidance scale to $\omega=2.0$, selected by ViSQOL on MUSDB18, and use it unchanged on both test sets. Both baselines use the splice of Sec.~\ref{ssec:synth}, whose effect is $<\!0.01$~ViSQOL. The comparison is nevertheless not strictly like-for-like: our predictor operates on quantized 12~kbit/s tokens, whereas the baselines use the raw low-band signal.

To assess whether the objective differences are perceptually meaningful, we conduct a MUSHRA test~\cite{itu2015bs1534} at the hardest setting, 8~kHz input, where three quarters of the output spectrum must be synthesized. We compare A2SB, the stronger external baseline overall, with our DAC predictor, which extracts the most from the frozen codec among our two models (Sec.~\ref{ssec:domains}). We select 12 MUSDB18 test excerpts: six from the half of the test set with the highest concentration of energy above 8~kHz, and six from the other half, covering both cases where a system must synthesize content and cases where restraint is rewarded. Each trial presents the labeled reference followed by four samples: the hidden 48~kHz reference, the band-limited input anchor, and the two systems. Ten non-expert listeners participated under standard office conditions.

% ===========================================================================
\section{Results}
\label{sec:results}
% ===========================================================================

\begin{table*}[t]
\caption{Objective metrics at every input rate, with the two corpora grouped: MUSDB18 (in domain) left, OrchideaSOL (out of domain) right. Both of our systems are the single multi-rate predictor, on the codec named. Bold marks the best of the four systems on the three metrics that can arbitrate here. Waveform $\ell_1$ ($10^{-2}$) and SI-SDR (dB) are given for completeness.}
\label{tab:musdb_osol}
\centering
\footnotesize
\begin{tabular}{ll ccccc ccccc}
\toprule
& & \multicolumn{5}{c}{\textbf{MUSDB18} (in domain)}
  & \multicolumn{5}{c}{\textbf{OrchideaSOL} (out of domain)}\\
\cmidrule(lr){3-7}\cmidrule(lr){8-12}
Input & System
& ViSQOL $\uparrow$ & Mel $\downarrow$ & STFT $\downarrow$ & Wav. $\downarrow$ & SI-SDR $\uparrow$
& ViSQOL $\uparrow$ & Mel $\downarrow$ & STFT $\downarrow$ & Wav. $\downarrow$ & SI-SDR $\uparrow$\\
\midrule
8\,kHz
& A2SB                 & 2.56 & 1.14 & 3.77 & 2.17 & 14.89 & 2.66 & 1.06 & 3.23 & 1.07 & 24.58\\
& UniverSR             & 2.77 & 0.93 & 2.73 & 2.25 & 14.23 & \textbf{2.91} & \textbf{0.83} & \textbf{2.49} & 1.48 & 24.91\\
& Ours (SpectroStream) & \textbf{2.99} & 0.77 & 2.49 & 2.49 & 14.01 & 2.71 & 0.98 & 2.68 & 1.16 & 26.73\\
& Ours (DAC)           & 2.88 & \textbf{0.76} & \textbf{2.46} & 2.28 & 14.66 & 2.89 & 0.90 & 2.56 & 1.20 & 26.51\\
\midrule
16\,kHz
& A2SB                 & 3.06 & 0.63 & 2.38 & 1.17 & 20.73 & 2.77 & 0.63 & 2.37 & 0.29 & 36.26\\
& UniverSR             & 2.41 & 1.23 & 3.33 & 0.99 & 22.14 & 2.80 & 0.77 & 2.54 & 0.24 & 38.27\\
& Ours (SpectroStream) & \textbf{3.40} & \textbf{0.42} & \textbf{1.77} & 1.21 & 20.36 & 3.00 & 0.46 & \textbf{1.84} & 0.30 & 37.70\\
& Ours (DAC)           & 3.28 & 0.45 & 1.83 & 1.17 & 20.78 & \textbf{3.11} & \textbf{0.43} & 1.86 & 0.32 & 37.67\\
\midrule
24\,kHz
& A2SB                 & 3.80 & 0.38 & 1.64 & 0.53 & 27.23 & 3.48 & 0.41 & 1.71 & 0.12 & 44.07\\
& UniverSR             & 3.16 & 0.74 & 2.41 & 0.47 & 28.11 & 3.36 & 0.48 & 1.98 & 0.11 & 44.20\\
& Ours (SpectroStream) & \textbf{3.94} & \textbf{0.27} & \textbf{1.36} & 0.62 & 25.98 & 3.54 & 0.27 & \textbf{1.42} & 0.14 & 43.05\\
& Ours (DAC)           & 3.90 & \textbf{0.27} & 1.37 & 0.59 & 26.45 & \textbf{3.64} & \textbf{0.25} & 1.44 & 0.14 & 43.11\\
\midrule
32\,kHz
& A2SB                 & 4.29 & 0.23 & 1.15 & 0.26 & 34.00 & 4.14 & 0.22 & \textbf{1.11} & 0.06 & 49.81\\
& UniverSR             & \multicolumn{5}{c}{\textit{rate not supported}} & \multicolumn{5}{c}{\textit{rate not supported}}\\
& Ours (SpectroStream) & 4.33 & \textbf{0.18} & \textbf{1.05} & 0.36 & 30.82 & 4.13 & 0.16 & \textbf{1.11} & 0.09 & 46.83\\
& Ours (DAC)           & \textbf{4.37} & 0.19 & \textbf{1.05} & 0.29 & 32.90 & \textbf{4.26} & \textbf{0.15} & 1.13 & 0.08 & 48.27\\
\bottomrule
\end{tabular}
\vspace{-10pt}
\end{table*}

\subsection{Comparative evaluation}
\label{ssec:quality}

On MUSDB18 (Table~\ref{tab:musdb_osol}) both of our predictors are ahead of both baselines at every input rate and on all three objective metrics. The ViSQOL margin over A2SB runs from more than $0.4$ at 8~kHz down to a few hundredths at 32~kHz, narrowing as the input widens and the task shrinks, and the multi-resolution mel and STFT distances order the four systems the same way.

Out of domain the ordering survives from 16~kHz upwards. On OrchideaSOL we lead on ViSQOL and mel at 16, 24 and 32~kHz and on STFT at 16 and 24~kHz, and only at 8~kHz does a baseline come out ahead, UniverSR on all three. UniverSR was trained on solo-instrument corpora close to that domain, absent from our own training data, so we generalize to unseen material and still lead at the wider inputs.

\noindent\textbf{Listening test.} Table~\ref{tab:mushra} reports the MUSHRA scores. The hidden reference is rated 100 at all three quartiles, and both systems sit far above the band-limited anchor. A2SB is rated slightly higher at every quartile, by 3 points at the median, but the interquartile ranges of the two systems overlap over most of their extent. Our predictor therefore performs comparably to A2SB at the hardest rate, although it reads quantized tokens at 12~kbit/s where A2SB reads the raw low band. The ViSQOL advantage of Table~\ref{tab:musdb_osol} at 8~kHz does not carry over to listeners, and listening to the predictions suggests why: our predictor fills the high band more completely than A2SB, which ViSQOL rewards, but at the cost of slight artifacts that listeners penalize.

\subsection{Discussing the results with the two different codecs}
\label{ssec:domains}

SpectroStream's tokens carry an explicit notion of frequency, and one may suspect that this is what makes bandwidth extension work in the token domain at all. DAC settles it: its tokens carry no frequency axis, yet the same recipe, held identical in architecture, cutoff draw, data, schedule, optimizer, band-limiting, synthesis and scoring, places it alongside SpectroStream throughout Table~\ref{tab:musdb_osol}. What the method exploits is the completion task itself, not a property of one codec.

The two are close but not equivalent. SpectroStream leads on ViSQOL in domain at 8, 16 and 24~kHz and DAC at 32~kHz, whereas DAC is the better of the two out of domain at every rate. They also do not leave the same room for improvement. Measured against what perfect prediction would give on each codec at the same bit rate, the DAC predictor captures $53.5$, $65.3$, $69.1$ and $69.3\%$ of the available headroom against $47.4$, $58.0$, $62.6$ and $57.6\%$ for SpectroStream, so SpectroStream's better in-domain scores reflect a higher ceiling rather than a token stream that is easier to complete. One factor could not be equalized and runs against DAC: fixing batch size and step count forced excerpt lengths of 2.5~s against SpectroStream's 5~s, and shorter excerpts cost quality, although DAC still sees the longer token sequence, 234 frames against 125.

\begin{table}[t]
\caption{MUSHRA scores (0--100) at 8~kHz input on MUSDB18, from 10 listeners over 12 excerpts, given as median and interquartile range.}
\label{tab:mushra}
\centering\footnotesize
\begin{tabular}{lcc}
\toprule
Condition & Median & $[Q_1, Q_3]$\\
\midrule
Hidden reference         & 100 & $[100, 100]$\\
Band-limited anchor      & 32  & $[18.75, 49]$\\
A2SB~\cite{kong2025a2sb} & 70  & $[59.75, 79.25]$\\
Ours (DAC)               & 67  & $[53.75, 76.25]$\\
\bottomrule
\end{tabular}
\vspace{-10pt}
\end{table}

\begin{table}[t]
\caption{ViSQOL on MUSDB18 across input rates, SpectroStream predictors. Per-rate: four predictors, one per rate. Multi-rate: the single predictor we propose. $+$rate emb.: the same, with a learned embedding of the cutoff. Anchor: the band-limited input scored as it stands, with nothing synthesized. Ceiling: the true full-band codes decoded at full depth, i.e.\ perfect prediction at the same bit rate.}
\label{tab:multirate}
\centering\footnotesize
\begin{tabular}{lccccc}
\toprule
Input & Anchor & Per-rate & \textbf{Multi-rate} & $+$rate emb. & Ceiling\\
\midrule
8\,kHz  & 1.59 & 3.01 & \textbf{2.99} & 2.98 & 4.55\\
16\,kHz & 1.92 & 3.43 & \textbf{3.40} & 3.39 & 4.48\\
24\,kHz & 2.98 & 3.93 & \textbf{3.94} & 3.93 & 4.51\\
32\,kHz & 4.01 & 4.35 & \textbf{4.33} & 4.30 & 4.56\\
\bottomrule
\end{tabular}
\vspace{-10pt}
\end{table}

\subsection{Predictor design}
\label{ssec:design}

Two decisions carry the multi-rate claim: training one predictor over all four input rates instead of one per rate, and giving it no indication of the input bandwidth. Table~\ref{tab:multirate} tests both, bracketed by a band-limited anchor and the codec's own ceiling. Every trained predictor sits well inside that interval, gaining $+1.40$, $+1.48$, $+0.96$ and $+0.31$ ViSQOL over the anchor as the input widens and less is left to synthesize. One predictor covers all four rates at a cost of at most $0.03$ ViSQOL against four dedicated ones, and at 24~kHz it is marginally ahead of them. Adding the cutoff embedding brings no benefit either: it sits below the multi-rate predictor at all four rates, by a few hundredths at most. The bandwidth is therefore already recoverable from the codes, and an explicit embedding supplies no information the predictor lacks.

\noindent\textbf{Where the bandwidth information is encoded.} That raises a question the framing does not answer: where in the codes does the model read the input bandwidth, and is it present at every RVQ level or only at level 1? It is not in individual latent dimensions. DAC's are not organized by frequency, and none of SpectroStream's 256 bottleneck dimensions survives band-limiting, which is expected once a learned projection merges its five latent frequency bands into that bottleneck. We therefore put the question to the codebooks themselves, asking how much the entry a level selects reveals about the frequency content of its input. Narrowband noise $400$~Hz wide is swept over 240 center frequencies from 100~Hz to 23~kHz, and we record which entry every RVQ level selects at each of them. Noise is preferred to a pure tone, whose spectral leakage varies with frequency and would itself resemble tuning. Two statistics summarize a level. The first asks how much a single token tells us about which part of the spectrum is active: the mutual information $I$ between the selected entry and the center frequency of the probe, corrected for small-sample bias by subtracting its value after shuffling the entries. The second asks whether the codebook is arranged by frequency, a property the first cannot see: the correlation $r$, over all pairs of entries, between their distance in codebook space and the distance between the center frequencies at which they are most often selected.

The two codecs use their level-1 codebook very differently. Under the probe, SpectroStream selects only 32 of its 1024 entries and its corrected mutual information is $1.82$ bits, while DAC selects 314 entries and reaches $3.04$ bits. DAC's level-1 token therefore localizes the active part of the spectrum more precisely, but spreads that information over an alphabet ten times larger. What separates them is arrangement rather than content, with $r=0.438$ for SpectroStream against $0.218$ for DAC. That gap is unchanged when both are restricted to their 32 most frequent entries, so it is not an artifact of alphabet size. The structure is also confined to level 1: at levels 2 to 4 the same measurement on SpectroStream reproduces it neither in sign nor in magnitude.

Two things follow, and a third does not. The bandwidth is legible in the level-1 codebook of both codecs, which is why no rate embedding is needed, and legible essentially nowhere deeper, which is why one decoded level suffices (Sec.~\ref{ssec:synth}). But the frequency ordering is not why the method works: DAC, whose level-1 codebook is the less ordered, is the codec whose predictor captures more of its headroom at every rate.

\iffalse
\begin{figure}[t]
\centering
\setlength{\fboxsep}{3pt}%
\fbox{\begin{minipage}{0.94\linewidth}\centering
\vspace{3mm}
{\scriptsize\textsc{placeholder: level-1 codebook frequency map}\\[2pt]
\textit{two panels, SpectroStream and DAC: entry selected by the narrowband}\\
\textit{probe (y, ordered by position in codebook space) against}\\
\textit{probe center frequency (x, log scale)}}
\vspace{3mm}
\end{minipage}}
\caption{Level-1 codebook under the narrowband-noise probe. Ordering the vertical axis by position in codebook space makes both statistics visible at once: the width of the cloud at a given frequency shows how precisely a token localizes the spectrum, its monotonicity whether the codebook is ordered by it.}
\label{fig:codebook}
\vspace{-10pt}
\end{figure}
\fi

% ===========================================================================
\section{Conclusion}
\label{sec:conclusion}
% ===========================================================================

Casting bandwidth extension as token completion puts the whole task on a decoder-side predictor and leaves the codec untouched. The cutoff frequency never has to be supplied, and probing the codebooks shows why: the input bandwidth is legible in the level-1 entries of both codecs and essentially nowhere deeper, which is also why decoding a single level is sufficient.

The proposed approach transfers unchanged between a spectral-domain and a waveform-domain codec, so it does not rely on tokens carrying a frequency axis, and while reading a 12~kbit/s token stream it exceeds A2SB and UniverSR on music mixtures at every rate. At 8~kHz, the hardest rate, listeners rate it close to A2SB, and out of domain it leads at every rate except that one. The predictor is still fitted to one frozen token distribution, so training a single predictor across several codecs, in the spirit of~\cite{li2024masksr,yang2024genhancer}, is the natural continuation.

\vfill\pagebreak

\bibliographystyle{IEEEbib}
\bibliography{refs_icassp}

\end{document}